\documentclass[aps, prx, reprint, longbibliography, nofootinbib,superscriptaddress,floatfix]{revtex4-2}
\pdfoutput=1

\usepackage{slashed}
\usepackage{verbatim}
\usepackage[T1]{fontenc}
\usepackage[utf8]{inputenc}
\usepackage[colorlinks]{hyperref}
\usepackage{mathbbol}
\usepackage[dvipsnames]{xcolor}
\usepackage[normalem]{ulem}
\usepackage{svg}
\usepackage[english]{babel}
\usepackage{lipsum}
\usepackage{physics}
\usepackage{dcolumn}
\usepackage{tensor}
\usepackage{comment}
\usepackage{graphicx,color,overpic,mathtools}
\usepackage{amsthm,amsmath,amssymb,hyperref,mathrsfs}
\usepackage{braket,bm,bbm,setspace}
\usepackage{cancel}
\usepackage{float}
\hypersetup{
    colorlinks=false,
    pdfborder={0 0 0},
}
\usepackage{xargs}
\newcommand{\ScriP}{\mathscr{I}^+}
\newcommandx*\wQNM[2][2=0]{\ensuremath{\omega_{#1 \, #2}}}

\newcommand{\ScalarProduct}[2]{\langle\langle #1 \, , #2 \, \rangle \rangle}
\def\Sh{\ensuremath{{}_s S_{\ell m \omega}}}
\def\Rh{\ensuremath{ {}_s R_{\ell m \omega }}}

\begin{document}

\title{Ringdown of a dynamical spacetime}

\author{Jaime Redondo--Yuste}
\email[]{jaime.redondo.yuste@nbi.ku.dk}
\affiliation{Niels Bohr International Academy, Niels Bohr Institute, Blegdamsvej 17, 2100 Copenhagen, Denmark}
\author{David Pere\~niguez}
\email[]{david.pereniguez@nbi.ku.dk}
\affiliation{Niels Bohr International Academy, Niels Bohr Institute, Blegdamsvej 17, 2100 Copenhagen, Denmark}
\author{Vitor Cardoso}
\affiliation{Niels Bohr International Academy, Niels Bohr Institute, Blegdamsvej 17, 2100 Copenhagen, Denmark}
\affiliation{CENTRA, Departamento de F\'{\i}sica, Instituto Superior T\'ecnico -- IST, Universidade de Lisboa -- UL,
Avenida Rovisco Pais 1, 1049-001 Lisboa, Portugal}
\affiliation{Yukawa Institute for Theoretical Physics, Kyoto University, Kyoto
606-8502, Japan}

\begin{abstract}
The gravitational waves emitted (some time) after two black holes merge are well described by the theory of linear perturbations on a spacetime characterized by the mass and spin of the remnant. However, in the very early stages right after merger, both the mass and spin are changing. In this work we explore, in a set up based on Vaidya's spacetime, the dynamical consequences of a change of mass in the spacetime due to the accretion of null matter (for example, gravitational waves). We show that accretion imprints time-dependent frequencies and amplitude to a ringdown waveform, and we show how to model accurately this effect in certain regimes. We also comment on the direct emission of gravitational waves due to perturbations in the in--falling matter, which is of relevance for black holes embedded in astrophysical environments.  
\end{abstract}

\maketitle

%%%%%%%%%%%%%%%%%%%%%%%%%%%%%%%%%%%%%%%%
\section{Introduction}
%%%%%%%%%%%%%%%%%%%%%%%%%%%%%%%%%%%%%%%%
The gravitational wave (GW) emission some time after the merger of two astrophysical compact objects is well described by the propagation of linear perturbations on the geometry of a single Kerr black hole (BH)~\cite{Anninos:1993zj,Berti:2009kk}. Such a \emph{ringdown} process is characterized by a superposition of damped sinusoids,  known as quasinormal modes (QNMs), whose complex quasinormal frequencies (QNFs) depend exclusively on the mass and spin of the BH~\cite{Vishveshwara:1970zz,Chandrasekhar:1975zza,Davis:1971gg, Kokkotas:1999bd, Berti:2009kk, Konoplya:2011qq}. {\it Black hole spectroscopy} is therefore a particularly interesting prospect for testing General Relativity (GR), since the detection of more than one QNM would allow to test GR predictions~\cite{Andersson:1997rn,  Dreyer:2003bv, Berti:2005ys}. 

To date, there are accurate waveforms capturing the ringdown~\cite{Baker:2008mj, Damour:2014yha, DelPozzo:2016kmd, Bohe:2016gbl, Cotesta:2018fcv, Nagar:2019wds, Nagar:2020pcj, Pompili:2023tna}. However, these model the early post-merger stage in a phenomenological way, by allowing a sufficiently large number of fitting parameters, thus not discerning the physical nature of the effects (non-linearities, changes in mass and or spin...). This makes it difficult to extend these models to more generic situations, as well as to perform targeted tests of GR, or even to generalize them for beyond GR theories. Thus, it is necessary to improve our physical understanding of the early post-merger phase in order to test GR accurately and minimize systematic errors~\cite{Moore:2021eok, Maggio:2022hre, Saini:2023rto, Purrer:2019jcp}. 

The regime of validity of linearized perturbations, i.e., of a waveform based on a superposition of damped sinusoids with constant frequencies and amplitudes, is restricted to several cycles after the merger~\cite{Dorband:2006gg, Berti:2007fi, London:2014cma, London:2018gaq, London:2018nxs, Bhagwat:2017tkm, Baibhav:2023clw}. Therefore, most of the analysis only consider a linear perturbation theory based model after $t \sim 10-15m$ after merger, with $m$ the final BH mass~\cite{LIGOScientific:2016lio, Carullo:2018sfu, Carullo:2019flw, Cotesta:2018fcv, Cotesta:2022pci, Pook-Kolb:2020jlr, Pook-Kolb:2020zhm}. The underlying reason is that the hypothesis that spacetime is a Kerr BH superposed with some \emph{small} perturbation is not correct in the early times after the merger. For instance, in~\cite{Bhagwat:2017tkm} it was shown that the ``Kerrness'' measures show significant deviations from the exact Kerr spacetime shortly after merger. In a similar vein, it was shown that the outer common horizon formed after the head--on collision of spinless BHs increases its area by several percent in the early ringdown stages~\cite{Mourier:2020mwa}. The fact that the remnant's mass and spin change significantly after the merger were already observed since the first numerical relativity breakthroughs~\cite{Buonanno:2006ui, Berti:2007fi}.

Delaying the starting time of the fit comes at a significant statistical cost, since the signal to noise ratio decays very quickly after merger. One way forward consists on understanding non--linear effects that could be important once the two BHs merge. Non--linearities have been shown to lead to turbulence, even for small perturbations of BHs in asymptotically anti--de--Sitter spacetimes~\cite{Adams:2013vsa, Bantilan:2012vu}, and there are some insights indicating that it could play a role also in the dynamics of asymptotically flat BHs~\cite{Yang:2014tla}. Non--linear effects have been identified also in the scattering of waves in BH spacetimes~\cite{Zlochower:2003yh, East:2013mfa} in head-on collisions of BHs~\cite{Gleiser:1996yc,Cheung:2022rbm}, and recently in the ringdown stage~\cite{London:2014cma, Cheung:2022rbm, Mitman:2022qdl}. There has been recent activity regarding modelling one such non--linear effect, which is the presence of quadratic QNMs (second order combinations of QNMs), which are the leading contribution in some higher harmonics for quasicircular binary coalescences ~\cite{Ripley:2020xby, Lagos:2022otp, Redondo-Yuste:2023seq, Perrone:2023jzq, Khera:2023lnc, Bucciotti:2023ets, Cheung:2023vki}. 

Given the violent merger process, leading to considerable luminosities in GWs, accounting for the changing mass and spin in the early stages of the ringdown is necessary in order to have waveform models that accurately describe the early post--merger stage. It turns out that a dynamically changing BH mass and spin can excite a plethora of modes~\cite{Sberna:2021eui} (see also ~\cite{Bamber:2021knr, Torres:2022fyf}). The analysis of these effects has been performed focusing on scalar fields in anti-de Sitter space, thus providing only a qualitative picture and guide for how would the mechanism work in asymptotically flat space, where the structure of infinity and the dynamics are drastically different. It is therefore important to include asymptotically flat BHs in the framework. We will significantly improve and extend previous studies, by considering a setup where backreaction due to absorption can be modelled exactly, at the cost of restricting to  configurations that are close to spherical symmetry.  

More precisely, we consider linear fluctuations of gravitational and pure radiation fields around Vaidya's spacetime, the latter being an exact solution to Einstein's equation describing accretion of radiation by a BH. This setup allows us to discuss in detail the effects of the mass change in the ringdown waveform. In particular, we provide a model that, by coupling the amplitudes, frequencies and phases to the evolving mass of the BH, is able to accurately capture the whole waveform, even in regimes where the timescale of the mass change is comparable to the oscillation frequency of the QNMs. An open question which we are also able to answer -- in the negative -- concerns possible echoes of GWs~\cite{Cardoso:2016rao,Cardoso:2016oxy}, caused by reflections off the infalling matter. We find no evidence for such a phenomena within our setup. Previous studies of perturbations on Vaidya spacetime~\cite{Abdalla:2006vb, Chirenti:2010iu} were restricted to scalar perturbations, which do not couple to neither the gravitational nor the pure radiation fluctuations. These studies already showed the coupling between the ringdown frequencies and the instantaneous mass of the BH. We extend them in several directions, by considering gravitational perturbations and uncovering the coupling to matter fluctuations, as well as by providing heuristic and accurate models of the ringdown signal on a Vaidya spacetime. 

The paper is organized as follows: first, in section~\ref{sec:Heuristics} we provide a simple heuristic argument that nevertheless captures some of the main features of ringdown in an accreting spacetime. In section~\ref{sec:Perturbation_Theory} we discuss the exact framework that we will be working with (gravitational perturbations in Vaidya spacetime), obtaining the master equation describing axial perturbations and discussing our numerical methods. We study our solutions in section~\ref{sec:results}, including extracting the mode content from the signal and finally proposing a novel waveform model based on time--varying amplitudes and frequencies, which captures better the waveform in the presence of accretion. We summarize our findings in Section~\ref{sec:Discussion}.

In the following we use geometric units $G = c = 1$, spacetime indices in $4$--dimensions are labelled with greek letteres $\mu,\nu,\dots = 0,\dots,3$, lower case roman letters are used to denote indices in the Lorentzian $2$--dimensional sphere sheaves $a,b = 0,1$, and upper case letters label coordinates in the $2$--sphere $A,B = 2,3$.

%%%%%%%%%%%%%%%%%%%%%%%%%%%%%%%%%%%%%%%%%%%%%%%%%%%
\section{Heuristics}\label{sec:Heuristics}
%%%%%%%%%%%%%%%%%%%%%%%%%%%%%%%%%%%%%%%%%%%%%%%%%%%

As a first exercise, we consider a simplified problem that may, however, illustrate some of the main features of BH relaxation in the presence of accretion. In particular, we will deal with the case of a very quick accretion process. The regime of very slow or adiabatic accretion presents additional difficulties due to the presence of possible secular effects, and we leave its study for future work.

Take a BH with initial mass $m_1$ relaxing in a ringdown process. The metric perturbation $h$ can be recovered uniquely (up to gauge redundancies~\cite{Chrzanowski:1975wv,Wald:1978vm}) from the Weyl scalar $\Psi^{[1]}_4 \equiv \Psi^{[1]}$, where the ``$[1]$'' superscript refers to the metric perturbation around the BH with mass $m_1$, and for simplicity we take it as a pure quadrupole QNM, with angular number $l = 2$.

Now consider that at some time, $t_1$, the BH undergoes an ``instantaneous'' mass increase $m_{1}\to m_{2}$ with $m_{2}>m_{1}$. Then, the evolution of the Weyl scalar at $t>t_{1}$, denoted by $\Psi^{[2]}$, is governed by the wave operators of the BH with mass $m_{2}$,  $\mathcal{O}^{[2]}_{s=-2} \Psi^{[2]} = 0$, subject to the initial conditions at $t_{1}$ given by a QNM of the BH with mass $m_{1}$, that is, $\Psi^{[1]}(t=t_1)$. At intermediate times $t > t_1$ the solution can be approximated by a superposition of QNMs $\psi^{[2]}_n$ of the larger BH~\footnote{
This way of writing the fluctuation is very suggestive for the ongoing discussion, but one should bear in mind it is only an approximation to the exact solution, that does not capture neither the direct emission nor the power--law tail contributions~\cite{Leaver:1986gd}.}
\begin{equation}
    \Psi^{[2]}(t,r) = \sum_{n=0}^\infty c^{[2]}_n e^{-i\omega^{[2]}_n (t-t_1)} \psi^{[2]}_n(r) \, 
\end{equation}
where $c^{[2]}_n$ are excitation factors to be determined from the initial conditions at $t = t_1$, $\omega_n^{[2]}$ are the QNFs of the BH with mass $m_2$ and $\psi^{[2]}_n$ are the radial wavefunctions of the QNMs with frequency $\omega_n^{[2]}$. In~\cite{Green:2022htq} (see also~\cite{Sberna:2021eui}) it was shown that such QNM excitation coefficients can be computed from the initial conditions by projecting them onto QNMs using a bilinear form, denoted $\ScalarProduct{\cdot}{\cdot}$, with respect to which QNMs with different frequencies are orthogonal. We provide a summarized discussion of this bilinear in Appendix~\ref{App:Bilinear}. In particular, if $\Psi^{[1]}(t=t_1)$ are the initial conditions, then we can write 
\begin{equation}
    c^{[2]}_n = \frac{\ScalarProduct{\Psi^{[1]}(t=t_1)}{\psi^{[2]}_n}}{\ScalarProduct{\psi^{[2]}_n}{\psi^{[2]}_n}} \, .
\end{equation}
Now if for simplicity we take the initial perturbation to consist only of the fundamental mode ($n = 0$), the solution at all times can be written as 
\begin{equation}
       \Psi(t) = A_0\begin{cases}
        e^{-i\omega^{[1]}_0 t}\psi^{[1]}_0 \, , \quad & t < t_1 \, , \\ 
        \sum_{n=0}^\infty e^{-i(\omega^{[1]}_0 -\omega^{[2]}_n)t_1} C_{0n} e^{-i\omega^{[2]}_n t}\psi^{[2]}_n \, , \quad & t > t_1 \, .
    \end{cases}
\end{equation}
Above we have defined the mixing coefficients $C_{kn}$,
\begin{equation}
    C_{kn} = \frac{\ScalarProduct{\psi^{[1]}_k}{\psi^{[2]}_n}}{\ScalarProduct{\psi^{[2]}_n}{\psi^{[2]}_n}} \, .
\end{equation}
Thus, at late times we can observe two distinct effects: (i) the off-diagonal terms $C_{0n}$ excite overtones of the final BH from the initial fundamental mode. This can be seen as an absorption induced mode excitation (AIME) effect~\cite{Sberna:2021eui}. (ii) The diagonal term, $C_{00}$, can be generically different than $1$. This results in a renormalization of the amplitude of the fundamental mode, a similar effect to that of second order perturbation theory~\cite{Lagos:2022otp}. We refer to this process as a \emph{decoherence} effect, since in practice the absorption is causing the initially coherent signal of a single mode to become a superposition of several modes, with different amplitudes. 

\begin{figure}
    \centering
    \includegraphics[width = 0.9\columnwidth]{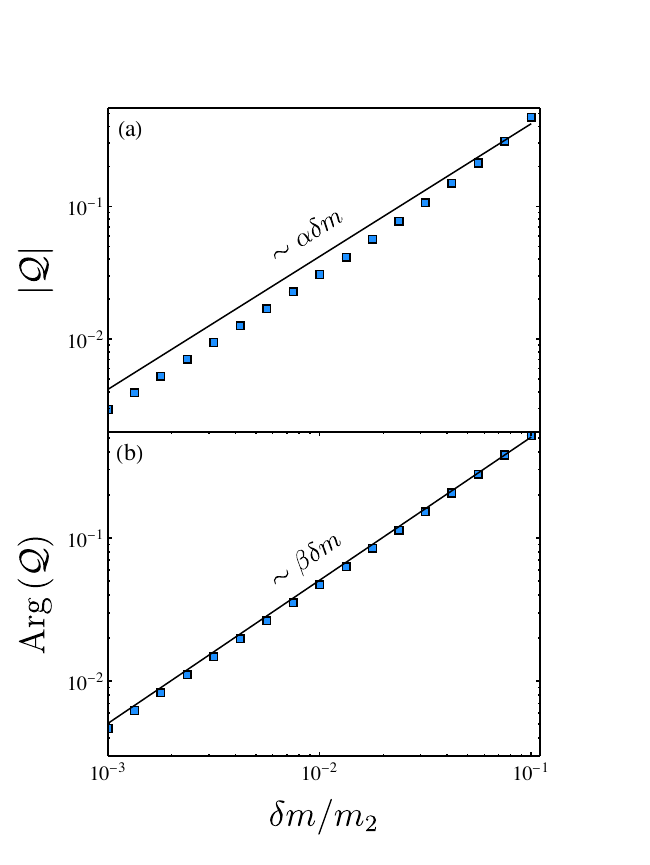}
    \caption{Decoherence factor $\mathcal{Q} = \lvert C_{00} -1 \rvert$ as a function of the relative change of mass $\delta m / m_2$. \textbf{Top: } Absolute value $|\mathcal{Q}|$. \textbf{Bottom: } Argument, defined as $\mathcal{Q} = |\mathcal{Q}| e^{-i\mathrm{Arg}(\mathcal{Q})}$. For comparison, we show the fit to a line of the form $\alpha \delta m$, where we find after fitting $m_2\alpha \sim 4.19 \pm 0.13$ and $m_2\beta = 5.07 \pm 0.03$.}
    \label{fig:Decoherence_Factor}
\end{figure}
We can then quantify which fraction of the initial QNMs amplitude projects into the fundamental mode of the BH after accretion, and which fraction is transmitted into certain number of overtones through AIME. We define the decoherence factor $\mathcal{Q}$ as
\begin{equation}\label{eq:Decoherence_Factor}
    \mathcal{Q} \equiv C_{00} - 1 =  \frac{\ScalarProduct{\psi^{[1]}_0}{\psi^{[2]}_0}}{\ScalarProduct{\psi^{[2]}_0}{\psi^{[2]}_0}} - 1 \, .
\end{equation}
This factor estimates how different is the amplitude that we would measure of the fundamental mode before and after accretion. Figure~\ref{fig:Decoherence_Factor} shows that the decoherence grows approximately linearly with the change of mass $\delta m = m_2-m_1$ (although higher order terms become important as $\delta m$ becomes larger), consistently with~\cite{Sberna:2021eui}. Moreover, the decoherence degree can be as large as $\sim 10\%$ for an accretion process where the mass changes only by a few percent. Although we will discuss in Section~\ref{sec:Discussion} in more detail the phenomenological implications of our work, this already suggests that absorption can have a large effect in the amplitude evolution of QNMs. 

The mass transition considered in this section is ``instantaneous'', and thus should only be regarded as an illustrative example. In what follows we consider instead a physically well-defined set up, where the effects of accretion on GWs are incorporated non-perturbatively.

%%%%%%%%%%%%%%%%%%%%%%%%%%%%%%%%%%%%%%%%
\section{Framework}\label{sec:Perturbation_Theory}
%%%%%%%%%%%%%%%%%%%%%%%%%%%%%%%%%%%%%%%%

The aim of this work is to explore the effects of accretion on the free oscillations of a BH. To do so we will consider fluctuations of exact solutions of GR that describe BHs accreting high frequency radiation, e.g. a GW. This set up allows us to account for the non-linear interactions between gravity and in-falling matter, while retaining the relative simplicity of perturbation theory.

%%%%%%%%%%%%%%%%%%%%%%%%%%%%%%%%%%%%%%%%
\subsection{Pure radiation fields and Vaidya spacetimes}
%%%%%%%%%%%%%%%%%%%%%%%%%%%%%%%%%%%%%%%%
A \textit{pure radiation field}, or \textit{null dust} \cite{Stephani:2003tm}, is a spacetime satisfying 
\begin{equation}\label{nulldust}
G_{\mu\nu}=\Phi K_{\mu}K_{\nu}\, , \ \ \ K^{\mu}K_{\mu}=0\, ,
\end{equation}
where $K^{\mu}$ and $\Phi$ are a null vector and a function. Conservation of the energy-momentum tensor (an immediate consequence of \eqref{nulldust}) implies that $K^{\mu}$ is geodesic, and without loss of generality, by simultaneous rescalings of $\Phi$ and $K^{\mu}$, it can be chosen to be affinely parametrised
\begin{equation}\label{nullgeod}
    K^{\mu}\nabla_{\mu}K^{\nu}=0\, .
\end{equation}
Physically, this is a spacetime describing the high-frequency (eikonal) approximation to unpolarized radiation, with energy density $\Phi$, propagating along the null direction $K^{\mu}$. This class of spacetimes was introduced by Vaidya \cite{Vaidya:1951zz}, and have proved useful in a wide range of physical scenarios ever since.

In double-null coordinates $(u,v,\theta,\phi)$, the spherically-symmetric line element reads~\cite{Girotto:2004iu}~\footnote{The factor $2$ is perhaps unconventional, but we choose to include it to follow the convention of~\cite{Girotto:2004iu, Abdalla:2006vb}. For the Schwarzschild geometry with mass $m$, $2f=(1-2m/r)$.}
\begin{equation}\label{linel}
    ds^{2}=-2f(u,v)dudv+r^{2}(u,v)d\Omega^{2}\,,
\end{equation}
which are well-defined coordinates as long as $f(u,v)\ne0$. Above, $r$ is the area radius function, and $d\Omega^2$ denotes the metric on the unit round $2$--sphere. Without loss of generality, $K^{\mu}$ can be taken to point along one of the null directions, say $K\sim \partial_{u}$. Then, up to rescalings of $K^{\mu}$ that depend on $v$ only, Eq.~\eqref{nullgeod} fixes
\begin{equation}\label{k}
    K=\frac{1}{ f(u,v)}\partial_{u}\, .
\end{equation}
The remaining equations do not fix the solution completely, and allow the free choice of a ``mass-profile function'' $m(v)$ (in terms of the Riemann tensor, $m=\frac{1}{2}r^{3}\tensor{R}{^\theta^\phi_\theta_\phi}$)\cite{Waugh:1986jh}. Restricting to mass profiles with $\partial_{v}m(v)\ne0$, Einstein's equations can be reduced to a transport equation for $r(u,v)$ along $\partial_{v}$ (the direction transverse to the null dust $K^{\mu}$),
\begin{equation}\label{flowR}
    \partial_{v}r=-\varepsilon \left(1-\frac{2 m(v)}{r}\right)\, , \ \ \ \ \ \ \varepsilon\equiv-\frac{\partial_{v}m(v)}{2\lvert \partial_{v}m(v)\rvert} \, ,
\end{equation}
which has a unique solution once an initial condition $r(u,v_{0})$ is prescribed. Then, the functions $f$ and $\Phi$ are\footnote{We have implicitly made some non-generic choices in reducing the equations of motion to \eqref{flowR} and \eqref{constraintsfphi} \cite{Waugh:1986jh}, but the class of solutions considered here are general enough for our purposes.}
\begin{equation}\label{constraintsfphi}
    \begin{aligned}
f&=2\varepsilon \partial_{u} r \, ,\\
\Phi&=-4\varepsilon\frac{\partial_{v}m}{r^{2}}=2\frac{\lvert \partial_{v}m\rvert }{r^{2}}\, .
\end{aligned}
\end{equation}
Notice that the second of these equations implies that the weak-energy condition holds automatically. We shall restrict to solutions with $f>0$ in the regime of validity of the double--null coordinates, and fix the time orientation by declaring $K^{\mu}$ in \eqref{k} to be future-oriented (so that $\partial_{u}$ and $\partial_{v}$ are future-oriented, too). Along the trajectories of $K^{\mu}$ the area-radius function $r(u,v)$ varies according to
\begin{equation}
K^{\mu}\nabla_{\mu}r=2\varepsilon\, .
\end{equation}
Therefore, increasing ($\varepsilon<0$) or decreasing ($\varepsilon>0$) profiles of $m(v)$ correspond to in-going or out-going pure-radiation fields, respectively. We will consider smooth mass profiles that interpolate between constant initial and final values
\begin{equation}
    m\left(v\rightarrow\pm\infty\right)=\begin{cases} m_{2}\\ m_{1}\end{cases}\, ,
\end{equation}
and will choose the asymptotic condition for the flow equation \eqref{flowR} of $r(u,v)$ as follows. At a slice $v=v_{\rm max}$, where $v_{\rm max}\gg1$ is taken large enough to achieve the condition 
\begin{equation}
    \Biggl\lvert \frac{m_{2}-m(v_{\rm max})}{m_{2}} \Biggr\rvert\ll1\,,
\end{equation}
we demand that $r(u,v_{\rm max})$ satisfies
\begin{align}\label{instant}
    r(u,v_{\rm max})&=\frac{m_{2}-m_{1}}{\lvert m_{2}-m_{1} \rvert}\frac{v_{\rm max}-u}{2} \\ 
    &-2m(v_{\rm max})\ln \left\lvert \frac{r(u,v_{\rm max})}{2m(v_{\rm max})} -1\right\rvert\, .
\end{align}
This choice allows us to interpret $(u,v)$, at large $v$, as the usual retarded and advanced times corresponding to the asymptotic state of the Vaidya background. The future event horizon and future null infinity are located at $u \to \infty$, $v \to \infty$, respectively.

Before considering a specific example, we notice that given an in-going pure-radiation solution associated to $r(u,v)$ with increasing mass profile $m(v)$ (i.e. where $f,\Phi,K...$ are constructed from $r(u,v)$ and $m(v)$ using the equations above), there is an out-going pure-radiation solution associated to $\tilde{r}(u,v)\equiv r(-u,-v)$ with decreasing mass profile $\tilde{m}(v)\equiv m(-v)$.

\begin{figure}
    \centering
    \includegraphics[width = 0.9\columnwidth]{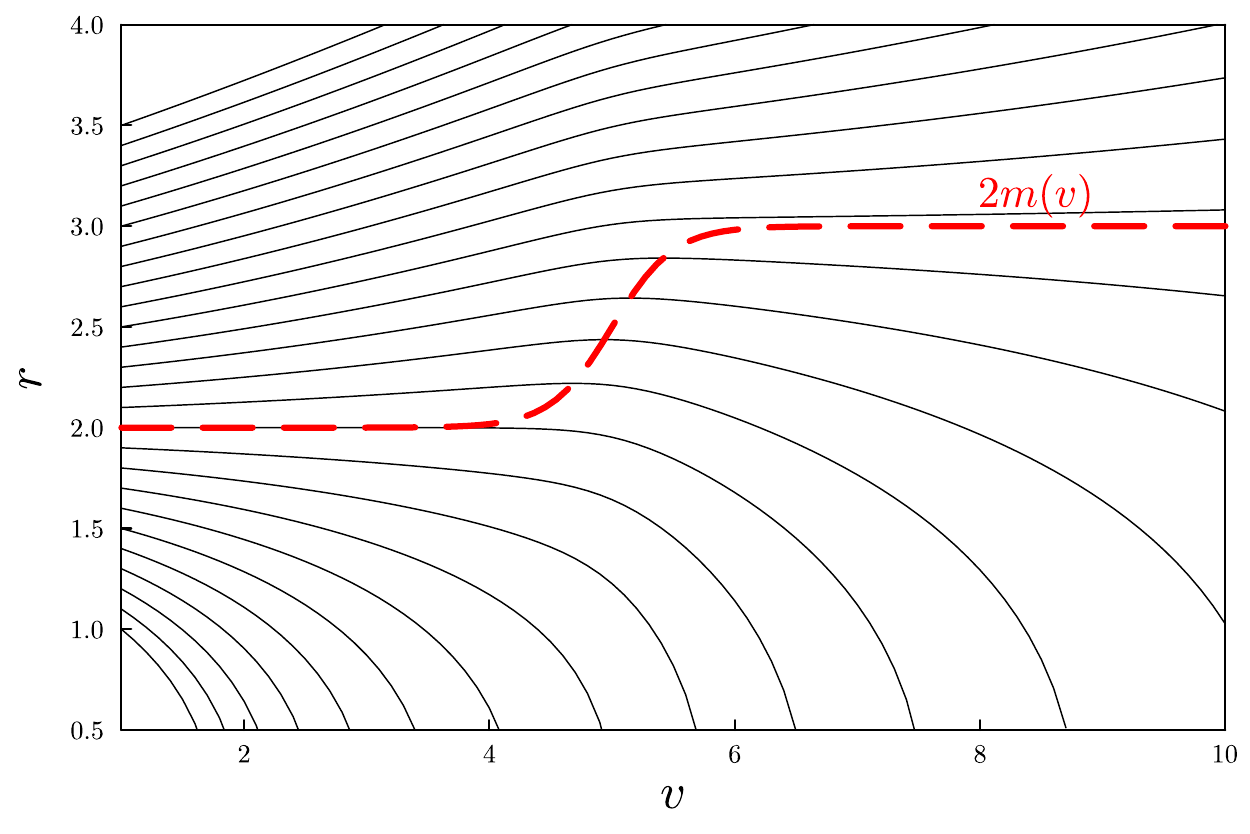}
    \caption{Lines of constant $u$ for the mass profile~\eqref{eq:mass_profile}, with $m_2 = 1.5m_1$, $v_1 = 5$, and $\tau = 0.5$. The red line shows the evolution of the mass profile $2m(v)$. Lines below this mass profile have always $r'<0$ and fall inevitably towards the singularity.}
    \label{Causal_Diagram}
\end{figure}
As a case study, consider a mass function of the form
\begin{equation}\label{eq:mass_profile}
    m(v) = m_1 + \frac{m_2-m_1}{2}\Bigl(1 + \tanh\bigl[\frac{v-v_1}{\tau}\bigr]\Bigr)\, ,
\end{equation}
where the mass increases from $m_1\to m_2$ in a time-scale controlled by the parameter $\tau$. This allows us to easily model different kinds of situations: ranging from the almost adiabatic increase of mass to very sudden changes. Integrating equation \eqref{flowR} numerically subject to condition \eqref{instant}, one finds a solution that describes a spherically symmetric BH increasing its mass from $m_{1}$ to $m_{2}$, as illustrated in the causal diagram in Fig.~\ref{Causal_Diagram}. 

%%%%%%%%%%%%%%%%%%%%%%%%%%%%%%%%%%%%%%%%
\subsection{Linear Fluctuations}
%%%%%%%%%%%%%%%%%%%%%%%%%%%%%%%%%%%%%%%%

Fluctuations of pure radiation fields consist of a gravitational perturbation $h_{\mu\nu}$ and the matter fields perturbations $\delta \Phi$ and $\delta K_{\mu}$, governed by equations \eqref{nulldust} linearised on the background introduced above. While our background is spherically-symmetric (and thus axial and polar sectors of the fluctuation decouple, see Appendix \ref{Ap1}), the fact that it is not vacuum entails a coupling between $h_{\mu\nu}$, $\delta \Phi$ and $\delta K_{\mu}$ that makes the analysis considerably more involved. Here we follow the covariant and gauge-invariant approach put forward in Ref.~\cite{Pereniguez:2023wxf} to handle spherically-symmetric background spacetimes with arbitrary matter content, which builds on previous groundbreaking work~\cite{Kodama:2000fa, Ishibashi:2003ap, Kodama:2003jz, Kodama:2003kk, Martel:2005ir,Chaverra:2012bh}. We shall simply report the master wave equations, and refer the interested reader to Appendix \ref{Ap1} and Ref.~\cite{Pereniguez:2023wxf} for details.

Being concerned mostly with free oscillations of BHs, in this work it will suffice to restrict to the axial sector of the fluctuation. As shown in Appendix \ref{Ap1}, after projection into spherical harmonics this sector is governed by two gauge-invariant master variables $\Psi(u,v)$ and $\tilde{\mathsf{v}}(u,v)$, encoding the gravitational and matter degrees of freedom respectively. These are subject to two coupled equations, following from the (linearised) Einstein's equation and the conservation of the energy-momentum tensor,\footnote{Here, lower case latin indices run from $0$ to 1, and all the geometric pieces (such as the volume form $\varepsilon_{ab}$ and covariant derivative $\nabla_{a}$) are associated to the metric $g_{ab}$ induced on surfaces of constant spherical angles (see Appendix \ref{Ap1}).} 
\begin{equation} \label{einVV}
\begin{aligned}
    r^{2}\nabla_{a}\left(r^{-2}\nabla^{a}(r\Psi)\right)-\frac{l(l+1)-2}{r}\Psi&=r^{2}\varepsilon^{ab}K_{b}\nabla_{a}\left(\Phi \tilde{\mathsf{v}}\right)\, , \\
     K^{a}\nabla_{a}\left(r^{2}\Phi \tilde{\mathsf{v}}\right)&=0 \, .
\end{aligned}
\end{equation}
In double-null coordinates, evaluating the above equations on the background implies that we can substitute the derivatives of $r(u,v)$ by
\begin{align}
\partial_{v}r&=-\varepsilon\left(1-\frac{2 m(v)}{r}\right)    \, , \\
\partial_{u}r&=\frac{f(u,v)}{2\varepsilon}\, ,\quad
\partial^{2}_{uv}r=-\frac{m(v) f(u,v)}{r^{2}}\, .
\end{align}
Remarkably, the second of Eqs.~\eqref{einVV} can be solved exactly for $ \tilde{\mathsf{v}}$, giving
\begin{equation}\label{genMat}
    \tilde{\mathsf{v}}=\frac{F(v)}{r^{2}\Phi}\,,
\end{equation}
where $F(v)$ is a free function of $v$, which corresponds to an initial condition for $ \tilde{\mathsf{v}}$.
In parallel, the first equation in \eqref{einVV} gives 
\begin{equation}
    \left[\partial^{2}_{uv}-\frac{f}{r}\left(\frac{3 m(v)}{r^{2}}-\frac{l(l+1)}{2 r}\right)\right]\Psi=\frac{2 f}{r^{2}}\varepsilon F(v)\, ,\label{eq:axial_master}
\end{equation}
which is the master wave equation we are seeking for. In a static background of mass $m$ one recovers the classic Regge-Wheeler equation (in particular, the only consistent choice in that case is $F(v)=0$, as follows from \eqref{genMat} and the fact that for constant $m$ one has $\Phi=0$).

Equations \eqref{genMat}--\eqref{eq:axial_master} are one of our main results. Matter perturbations are completely determined once $F(v)$ is prescribed and do not depend on the gravitational sector $\Psi$. Axial GWs, on the other hand, are sourced by matter fluctuations in a very simple way. We can contrast this behavior to axial perturbations of perfect fluids in stationary backgrounds~\cite{Chandrasekhar:1991fi, thorne1967non, Bamber:2021knr,Cardoso:2021wlq,Cardoso:2022whc}, where the gravitational fluctuation satisfies an homogeneous equation. In the other extreme are {\it polar} fluctuations (not considered here) which couple efficiently to the matter sector even in static spacetimes~\cite{Cardoso:2021wlq,Cardoso:2022whc}. Thus, results \eqref{genMat}--\eqref{eq:axial_master} are an interesting outcome of dealing with a special type of matter, null radiation, in a non-stationary background.

%%%%%%%%%%%%%%%%%%%%%%%%%%%%%%%%%%%%%%%%%%%%%%%%%%
\subsection{Numerical framework}\label{sec:Numerics}
%%%%%%%%%%%%%%%%%%%%%%%%%%%%%%%%%%%%%%%%%%%%%%%%%

%
\begin{figure}
    \centering
    \includegraphics[width = 0.9\columnwidth]{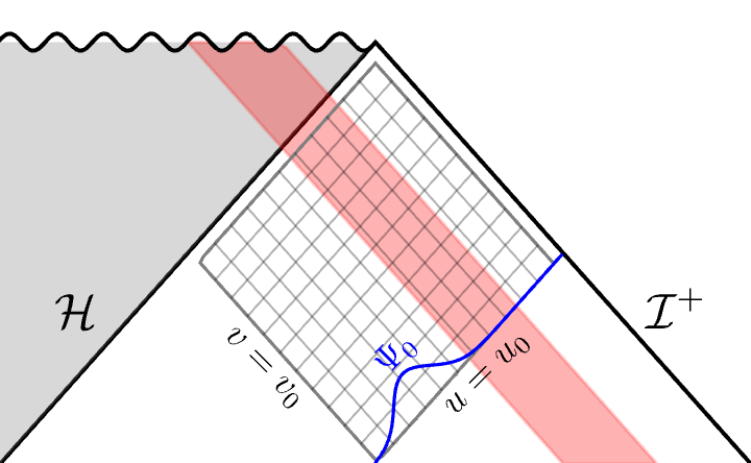}
    \caption{Upper half of the Penrose diagram describing Vaidya spacetime. The horizon grows due to absorption of null matter falling (red shaded region). Here, $\{u_0,v_0\}$ represent the region where we impose the initial conditions (blue curve), and $\{u_{\rm max},v_{\rm max}\}$ control how close does our grid get to the horizon and null infinity, respectively.}
    \label{fig:Diagram}
\end{figure}
We will solve the previous master equation numerically, making use of a characteristic algorithm. Here, we present the details of the numerical evolution, as well as the class of initial conditions that we will consider. The master equation that we are dealing with has the general form
\begin{equation}\label{eq:Numerics_Master_Equation}
\partial^2_{uv}\Psi + V(u,v)\Psi = \mathcal{S}[F] \, ,
\end{equation}
where $\Psi $ is the master variable, $V(u,v)$ is the potential and $\mathcal{S}$ is a source term that depends on the matter fluctuations $F$. We consider a finite discrete grid in the range $[u_0,u_{\rm max}]\times [v_0,v_{\rm max}]$, as shown in Fig.~\ref{fig:Diagram}. 

In order to prescribe initial conditions, we assume that the $v=v_0$ surface is located far enough that we can set $\Psi(u,v_0)=0$, giving initial data only in the $\Psi(u_0,v)$ surface. We consider a Gaussian wave packet
\begin{equation}\label{eq:Numerics_Initial_Condition}
    \Psi(u_0,v) = \Psi_0 \exp\Bigl(- \frac{(v-v_\Psi)^2}{\sigma_\Psi^2}\Bigr) \, .
\end{equation}
Whenever we consider non--vanishing matter fluctuations, the profile $F(v)$ will also be given by a Gaussian of the form~\eqref{eq:Numerics_Initial_Condition}. 

The evolution proceeds through the usual characteristic algorithm~\cite{Gundlach:1993tp}. We integrate the equations from left to right at each new constant $v$ slice. The first order equations of the background~\eqref{flowR}--\eqref{constraintsfphi} are solved using a fourth order finite difference approximation. The time update for the wave equation is given by 
\begin{equation}\label{eq:Numerics_Update_Rule}
    \Psi_N = \Psi_E + \Psi_W - \Psi_S + \frac{\Delta_u \Delta_v}{2} V_S (\Psi_E + \Psi_W) - \mathcal{S}_S \, ,
\end{equation}
where $N,E,S,W$ denote the north, east, south and west points as seen from the Penrose diagram, and $\Delta_u, \Delta_v$ are the grid resolutions in the $u$ and $v$ directions. A challenging point of the evolution appears when solving the radial equation Eq.~\eqref{flowR}, which needs instant conditions given by Eq.~\eqref{instant}. In order to efficiently solve for the initial conditions, we combine two different algorithms, depending on whether $r_\star = (v-u)/2$ is large enough~\cite{Gundlach:1993tp}. When $r_\star$ is large, a simple root--finding algorithm converges quickly. When $r_\star \lesssim 0$, 
we iterate the equation in its form $r / 2m_2 = 1 + \mathrm{exp}[(r_\star - r)/2m_2]$ until the desired accuracy is achieved. We denote the solution close to the horizon as $\Psi_\mathcal{H} \sim \Psi(u_{\rm max}, v)$, and at infinity $\Psi_\ScriP \sim \Psi(u, v_{\rm max})$. We have tested the accuracy of our algorithm by recovering the Schwarzschild QNFs to good accuracy, as well as Price's law tail exponents, as will be discussed below. Moreover, our code shows the expected convergence rate when increasing resolution.
The code is implemented in \texttt{Julia} and is made available through the \texttt{VaidyaPT.jl} repository~\cite{VaidyaPT}.

%%%%%%%%%%%%%%%%%%%%%%%%%%%%%%%%%%%%%%%%%%%%%%%%%%%%%%%%%%%%%%
\section{Results}\label{sec:results}
%%%%%%%%%%%%%%%%%%%%%%%%%%%%%%%%%%%%%%%%%%%%%%%%%%%%%%%%%%%%%

Once we have set up our problem and described the numerical methods employed to solve it, we move towards our goal of understanding ringdown processes in an accreting spacetime. First, we discuss the general features of the problem, including the excitation of a ringdown process by matter fluctuations. Secondly, we attempt to extract the mode characteristics on short timescales. Finally, we introduced a novel model, adapted to dynamical scenarios, and show convincing evidence that it outperforms the usual damped sinusoids framework. 

%%%%%%%%%%%%%%%%%%%%%%%%%%%%%%%%%%%%%%%%%%%%%%%%%%%%%%%%%%%%%%
\subsection{General Features}\label{sec:general_results}
%%%%%%%%%%%%%%%%%%%%%%%%%%%%%%%%%%%%%%%%%%%%%%%%%%%%%%%%%%%%%
%
\begin{figure*}
    \centering
    \includegraphics[width = 1.05\textwidth]{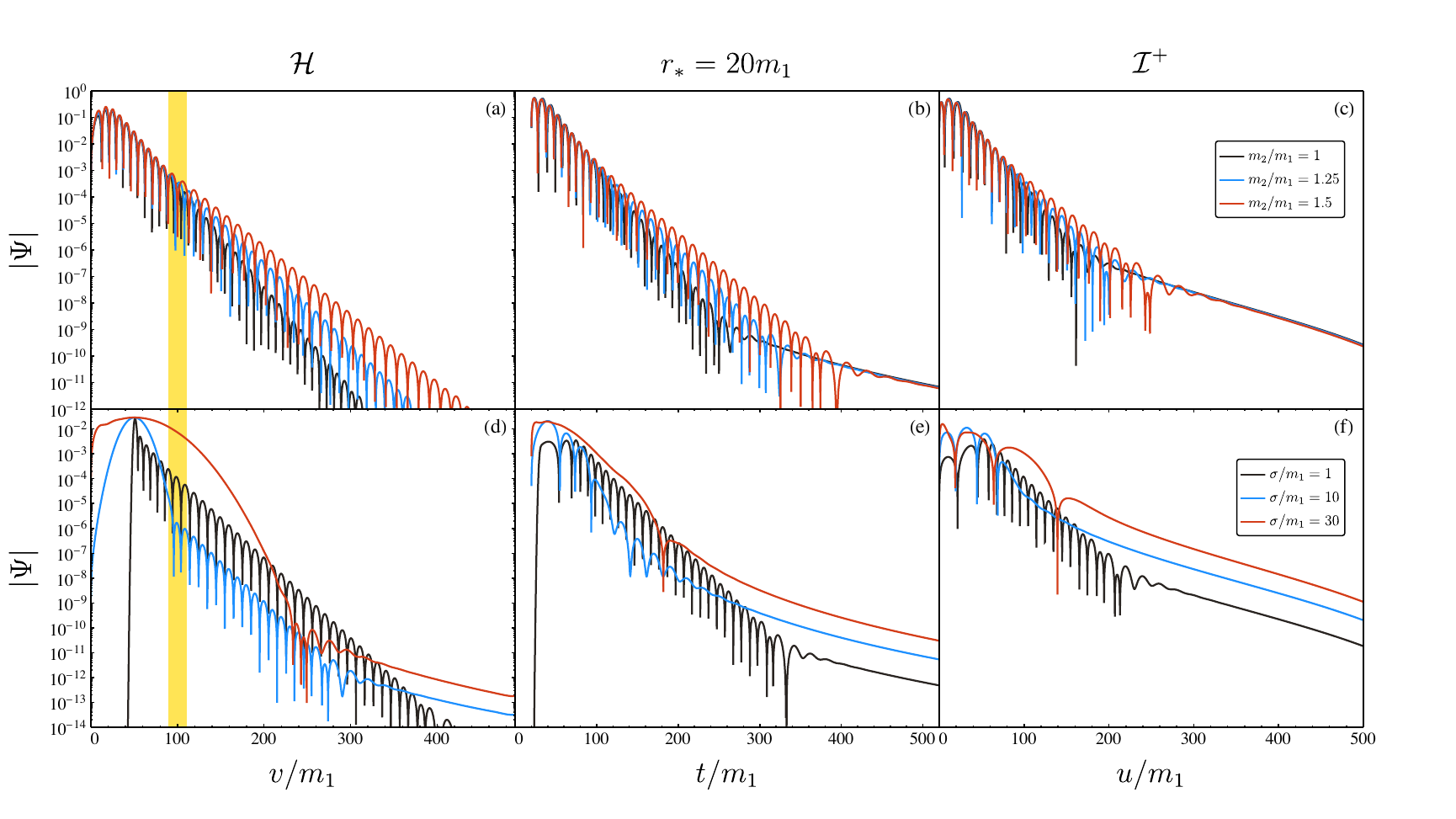}
    \caption{\textbf{Top: } Evolution of the quadrupolar gravitational master variable $\Psi$ extracted at \textbf{a.} the horizon, \textbf{b.} a fixed radius $r_*=20m_1$, and \textbf{c.} null infinity. Matter fluctuations vanish, while initial conditions are given by Eq.~\eqref{eq:Numerics_Initial_Condition} with $(\Psi_0,v_\Psi/m_1,\sigma_\Psi/m_1) = (1, 15, 2.5)$, evolving on a Vaidya background~\eqref{eq:mass_profile} with $(v_1/m_1,\tau/m_1) = (100,10)$ and $m_2/m_1$ as indicated in the legend. The shaded yellow region, thus, represents the region where the background is most dynamical 
    \textbf{Bottom: }Evolution of $\Psi$ when the initial conditions are trivial, $\Psi_0 = 0$, but there are matter fluctuations present. The profile for the matter fluctuations $F(v)$ is given by a Gaussian pulse, as in Eq.~\eqref{eq:Numerics_Initial_Condition}, with amplitude $0.01$, centered at $v/m_1 = 50$ and width $\sigma$. In this case, $m_2 = 1.2 m_1$. 
    }
    \label{fig:Example_Evolution}
\end{figure*}
The evolution of the master variable $\Psi$ (directly related to the GW strain $h$ via~\eqref{eq:Psi_to_Strain}), is shown in Fig.~\ref{fig:Example_Evolution}, for a background~\eqref{eq:mass_profile} with $(v_1/m_1,\tau/m_1) = (100,10)$ and different intensity of accretion, as measured by $m_2/m_1$. In the following, we focus on the dominant quadrupolar $l=2$ mode.
For $m_2=m_1$ we are simply describing the dynamics of slightly disturbed vacuum Schwarzschild BHs. For $m_2/m_1=1.5$ on the other hand, we are discussing a violently accreting spacetime, which saw a 50\% increase in its mass on a very short timescale (roughly 5 light-crossing times).

Consider first a sourceless evolution, where matter fluctuations are set to zero ($F$ in Eq.~\eqref{eq:axial_master} vanishes).
The top panels show the evolution for different values of the final mass $m_2$, with the black line corresponding to the evolution on a purely Schwarzschild background $m_2=m_1$. This relaxation process is known as quasinormal ringdown and can be understood as leakage from the light ring~\cite{Goebel:1972abc,Mashhoon:1985cya,stewart1989solutions, Ferrari:1984zz,Berti:2005ys,Berti:2009kk,Dolan:2009nk,Cardoso:2019rvt}. At early times it is the same for all backgrounds, since all backgrounds have same $m_1$; thus the light ring properties are identical early on. During this stage,
\begin{equation}
\Psi \sim e^{-i\omega_{\rm QNM} t}= e^{-t/\tau_{\rm QNM}}\cos (\omega^R_{\rm QNM} t +\phi)\,,\label{eq:def_qnms}
\end{equation}
where we assumed that there is a dominant QNM frequency which we write as
\begin{equation}
\omega_{\rm QNM}=\omega^R_{QNM}-i/\tau_{\rm QNM}\,.
\end{equation}
It is instructive to note that, for a fixed-mass, Schwarzschild spacetime of mass $m$, then 
\begin{equation}
m\omega^R_{\rm QNM}=0.373672\,, \qquad m/\tau_{\rm QNM}=0.0889623\,.\label{eq:modes_BH}
\end{equation}
One might therefore expect that early relaxation is described as above with $m \to m_1$, an expectation which is consistent with our results.

However, once accretion starts, the region near the light ring changes and so does the relaxation of the spacetime. Spacetimes with larger mass relax with a lower frequency and over longer timescales. Indeed, from Fig.~\ref{fig:Example_Evolution}, we observe that at early and late times the behaviour is as in Eq.~\eqref{eq:modes_BH} with $m\to m_{1,2}$, respectively, with a transient in between. We will explore this in more detail below. Notice, also, that the relaxation changes both in the waveforms extracted at the horizon (panel \textbf{a}) and in the waveform extracted at future null infinity (panel \textbf{c}). The frequency change as seen from $\ScriP$ is a direct consequence of the fact that the light ring grows.

We also see power-law tails at very late times, when the field decays as $\Psi\sim t^{-p}$~\cite{Price:1971fb,Leaver:1986gd,Ching:1994bd,Ching:1995tj}
at fixed radius and at null infinity. Similar power law tails form at the horizon, but only at much later times, for the class of initial conditions considered here. The exponent of the tail in this case is in agreement with Price's law~\cite{Price:1971fb}, ($t^{-7}$ for this case) when extracting at a fixed radius (panels \textbf{b-e}). Despite the spacetime being dynamical, it asymptotes towards a Scwharzschild background, hence we expect Price's law to be satisfied~\cite{Gundlach:1993tp, Hod:2002gb}.

The bottom panels of Fig.~\ref{fig:Example_Evolution} show the excitation of gravitational perturbations due to a matter fluctuation. We consider a matter profile for $F(v)$ given by a Gaussian pulse analogue to~\eqref{eq:Numerics_Initial_Condition}, centered at $v_1=50m_1$, with fixed amplitude and vary the width $\sigma$ (see legend in panel \textbf{f}). We find that very thin profiles excite a larger overtone contribution, as can be seen from the rapid decay at early times at the horizon, whereas wider pulses excite the tail more efficiently, as is clearly seen in panels \textbf{d--f}, consistently with~\cite{Berti:2006wq}. It is important to remark that in this case the matter fluctuations satisfy a first-order evolution equation that could be solved analytically, so there are no ``matter'' modes. As a consequence, the evolution of the gravitational perturbations $\Psi$ is qualitatively equivalent to having a non--zero initial condition for $\Psi$.  

%%%%%%%%%%%%%%%%%%%%%%%%%%%%%%%%%%%%%%%%%%%%%%%%%%%%%%%%%%%%%%%%%%%%%%
\subsection{Mode content of a dynamical ringdown}\label{sec:Modelling}
%%%%%%%%%%%%%%%%%%%%%%%%%%%%%%%%%%%%%%%%%%%%%%%%%%%%%%%%%%%%%%%%%%%%%%
\begin{figure*}
    \centering
    \includegraphics[width = 1.05\textwidth]{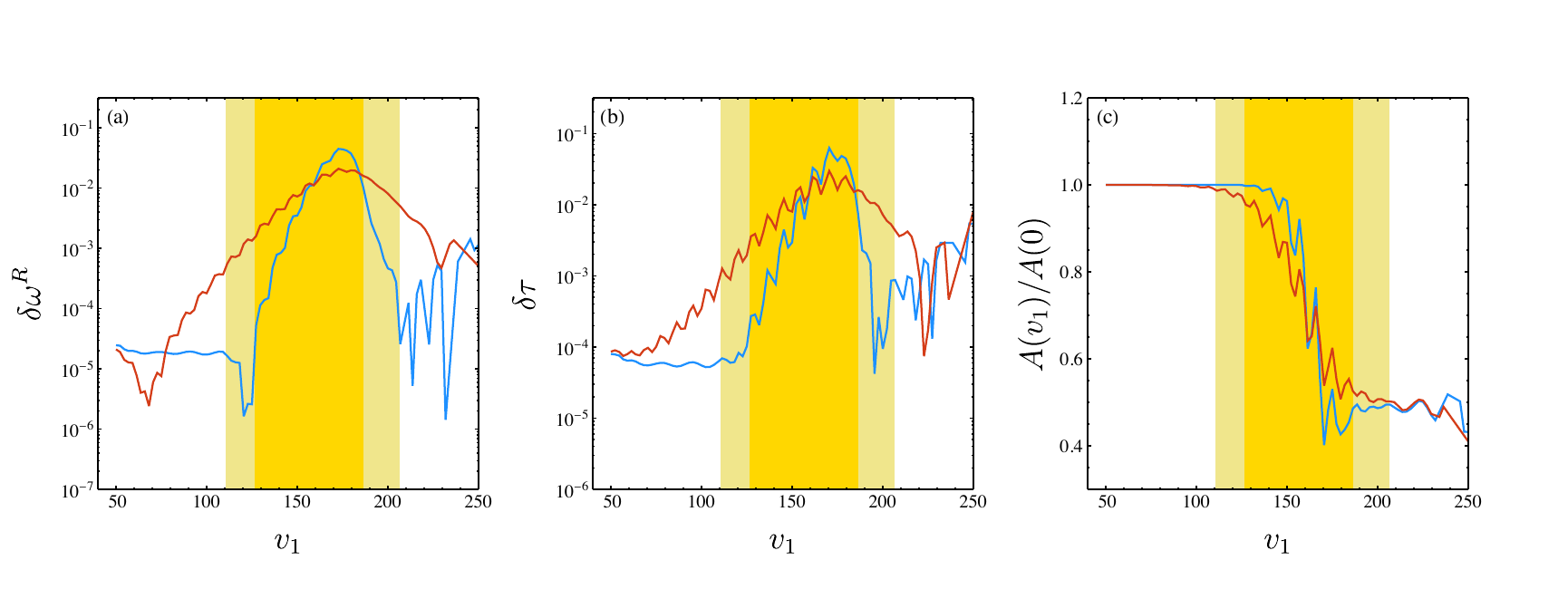}
    \caption{In all panels, we fit the waveform obtained for a Vaidya background with $m_2/m_1 = 1.05$ and $\tau = 5 (25)$ in blue (red), with a template containing one free damped sinusoid, at a time starting at $v = v_1$ (referred to the peak of the waveform), for $6$ half--periods. \textbf{a}. Relative error in the frequency $\delta f$ compared to the Schwarzschild value, see Eq.~\eqref{eq:Definition_Error_Freqs}. \textbf{b}. Same, but for the damping time. \textbf{c}. Extracted amplitude, normalized with respect to the initial amplitude. The shaded regions represent starting times such that the fitting window overlaps with the region where spacetime is most dynamical. Oscillations of the extracted values at late times are most likely due to contamination with numerical noise of the signal.}
    \label{fig:Instantaneous_Fits}
\end{figure*}
Once we have numerical evolutions of gravitational perturbations in the Vaidya background, we consider the problem of modelling those waveforms. Since the matter fluctuations do not add additional dynamical modes, for simplicity we consider only the case where the matter fluctuations vanish (therefore, the equation is homogeneous), and the gravitational perturbations are sourced by some initial conditions. Moreover, the frequency content should be independent on whether the perturbations are extracted at infinity or at the horizon. Since the matter that is accreted into the BH falls along the $\partial_u$ null direction, studying the waveform at the horizon (which is transverse to $\partial_{u}$) provides a cleaner picture. Thus, in the following we will study the metric perturbation at the horizon, $\Psi \equiv \Psi_{\mathcal{H}}$.

Ultimately we are describing a ringdown process, therefore, the fundamental ingredient of the waveform model used to describe the signal is expected be a combination of damped sinusoids, 
\begin{equation}\label{eq:Damped_Sinuosoids}
    \Psi = \sum_n A_n \cos(\omega^R_n v + \phi_n) e^{-v / \tau_n} \, , 
\end{equation}
where $A_n$ are the amplitudes, $\phi_n$ the phases, $\omega^R_n$ the oscillation frequencies and $\tau_n$ the damping times (see definition~\eqref{eq:def_qnms}). There is an ambiguity in defining the amplitudes and phases, since shifting the zero of the time $v \mapsto v - v_0$ rescales their values. We fix that ambiguity by re--scaling always the time axis so that $v = 0$ corresponds to the peak of the waveform. Therefore, amplitudes and phases reported here are always referred to the peak of the signal.

We employ a least squares fitting algorithm~\cite{liu1989limited, fletcher2000practical}, with performance boosted via automatic differentiation. We choose initial conditions for the algorithm by sampling from uniform priors in the ranges $A \in [0,1]$, $\phi\in[0,2\pi]$, $m_1\omega^R\in[0,1]$ and $\tau/m_1\in[1,20]$. We iterate on the algorithm until the mismatch between the reconstructed waveform $\Psi_R$ (i.e. the waveform evaluated at the best fit parameters) and the numerical signal is below a certain threshold.
We have tested our algorithm by simulating mock data and extracting accurately the parameters, even in the presence of Gaussian noise. 

By looking at Fig.~\ref{fig:Example_Evolution}-\textbf{a}, we can already observe that the waveform changes behaviour during the transient, in a way that depends on the value of the change of mass $\delta m = m_2-m_1$. When the mass changes, the oscillation frequencies and the damping time change. This is to be expected: intuitively the behaviour at early times should be governed by some combination of QNMs of the BH with mass $m_1$, and at late times the same should be true for a BH with mass $m_2$: hence, the dimensionless quantity $\omega^R\, m(v)$ should be the same at early and late times, although it may oscillate in the transient. 

In order to test this, we extract the average frequency over a time span $[v_1, v_1+v_N]$, where $v_N$ is chosen such that the signal contains approximately $N$ half-periods, and we choose $N=6$~\footnote{This was the smaller value of $N$ capable of recovering accurate enough the frequencies in tests containing Schwarzschild waveforms.}. This serves as an estimate for the ``instantaneous'' frequency, amplitude, and phase of the signal. We do so using different starting times, and show our results in Fig.~\ref{fig:Instantaneous_Fits}. We use only one damped sinusoid in these fits. 

First, we define the frequency and damping time shifts
\begin{equation}\label{eq:Definition_Error_Freqs}
        \delta \omega^R =\Bigl \vert 1 - \frac{m(v) \omega^R}{m_1 \omega^R_{\rm QNM}} \Bigr \vert \, , \quad\delta \tau = \Bigl \vert 1 - \frac{m_1 \tau}{m(v) \tau_{\rm QNM}} \Bigr \vert \, , 
\end{equation}
where $\omega^R$ and $\tau$ are the extracted oscillation frequency and damping time, and $m_1\omega^R_{\rm QNM},\,m_1\tau_{\rm QNM}$ are their QNM values for Schwarzschild with mass $m_1$. Since we are re--scaling by the appropriate mass dimension, we expect, as Fig.~\ref{fig:Instantaneous_Fits} shows, that these errors go to zero as $v_1 \to 0, \infty$. As one could naturally expect, when the mass absorption occurs more adiabatically (red line), the error in the frequency extraction is larger for a larger fraction of the evolution. On the other hand, when the mass absorption occurs very abruptly, the frequency errors $\delta \omega^R$ and $\delta \tau$ are quite small (of the order of $\sim 10^{-4}$ or below, notice that the scale is logarithmic), except for a very short time, related to the time when the background is dynamical, where they become larger than for their adiabatic counterpart. 

The right panel in Fig.~\ref{fig:Instantaneous_Fits} shows the evolution of the amplitude of the fundamental mode, $A(v_1) \equiv A_0$, where the fit occurs over $6$ half--periods starting at $v_1$. For convenience we normalize the amplitude with respect to the amplitude measured at early times (but once it is compatible with a constant), which we label as $A(0)$. We observe that at early times it is constant, which hints that a single damped sinusoid is providing an accurate representation of the signal. It is only at very late times that it becomes constant again, but the final value of the amplitude is substantially smaller than the initial one. Remarkably, it does \emph{not} depend on the timescale at which the background changes, $\tau$. 

We remark that at all times the mismatch between the reconstructed waveform and the numerical data is always below $\sim 10^{-3}$. The mismatch achieves its largest values whenever the fitting window overlaps with the timescale where the BH mass is growing more significantly.

\begin{figure}
    \centering
    \includegraphics[width = 0.95\columnwidth]{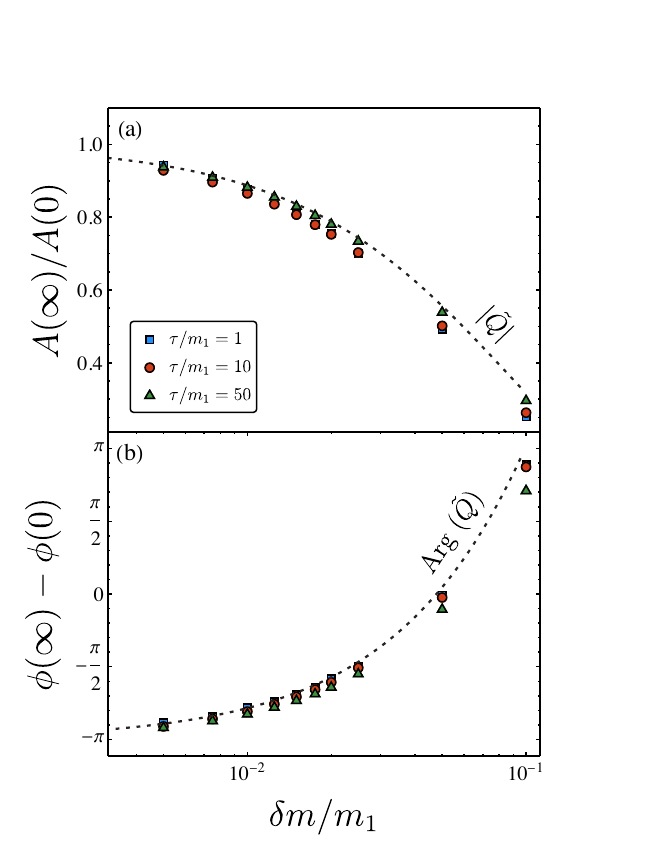}
    \caption{\textbf{Top:} Ratio between the final amplitude, $A(\infty)$, extracted using $6$ half--cycles starting at $v \geq v_1 + 4\tau$, and the initial amplitude, extracted using $6$ cycles starting at $v = 50$, for different values of the background profile $\{m_2,\tau\}$, as indicated in the legend. The dashed black line represents the (absolute value of the) re--scaled decoherence factor $\Tilde{\mathcal{Q}}$, as defined in Eq.~\eqref{eq:Decoherence_Factor_Rescaled}. In order to evaluate it, we use the linear fit extracted from Fig.~\ref{fig:Decoherence_Factor}. \textbf{Bottom:} Same, but for the phase difference. In this case the black dashed line represents the complex argument of the decoherence factor.}
    \label{fig:Amp_Ratio}
\end{figure}
Taking one step further towards modelling the signal, we extract directly the final (initial) amplitudes and phases, $A(\infty)$ ($A(0)$), $\phi(\infty)$  ($\phi(0)$), in the regime where they are approximately constant. If the discussion in Sec.~\ref{sec:Heuristics} is a good approximation, these should be related directly to the QNM decoherence factor $\mathcal{Q}$, as defined in Eq.~\eqref{eq:Decoherence_Factor}. However, that factor relates the amplitudes as referred to the time at which the mass changes (that time is given by $v_1$ in the mass profile~\eqref{eq:mass_profile}). On the other hand, the amplitudes and phases extracted from the fits are referred to the peak of the waveform, $v_{\rm peak}$. Thus, we would expect a relation given by
\begin{equation}\label{eq:Decoherence_Factor_Rescaled}
    \frac{\Tilde{A}(\infty)}{\Tilde{A}(0)} = (1+\mathcal{Q}) \frac{e^{i\omega(\infty) (v_1-v_{\rm peak})}}{e^{i\omega(0)(v_1-v_{\rm peak})}} \equiv \Tilde{\mathcal{Q}} \, , 
\end{equation}
where $\Tilde{A} = A e^{i\phi}$ is the complex amplitude, and $\omega(\infty)$ (resp. $\omega(0)$) are the final (initial) ringdown frequencies of the fundamental mode. Fig.~\ref{fig:Amp_Ratio} shows the extracted values for the absolute value and the phase from three different regimes: $\tau = 1,10,50$, ranging from non--adiabatic, to a more adiabatic situation, as a function of the mass increase $\delta m = m_2-m_1$. Our extracted values from the fit agree surprisingly well with the rescaled decoherence factor $\tilde{\mathcal{Q}}$. The agreement is remarkably good for all values of $\tau$ considered. On the other hand, the largest value of $\tau$ considered is only comparable to a few decades of decay of the fundamental mode frequency, and thus it is not yet exploring the really adiabatic regime. Pushing beyond this regime would require specifically targeted numerical methods that go beyond the scope of this work. 

%%%%%%%%%%%%%%%%%%%%%%%%%%%%%%%%%
\subsection{A dynamical ringdown model}
%%%%%%%%%%%%%%%%%%%%%%%%%%%%%%%%%
The previous analysis motivates a template that goes beyond damped sinusoids with fixed amplitudes and frequencies. Having knowledge of the mass function $m(v)$, and some analytical control as given by the decoherence factor $\mathcal{Q} \sim \alpha \delta m e^{-i\beta\delta m}$ (with $\alpha \sim 4.19$ and $\beta \sim 5.03$), as discussed in Fig.~\ref{fig:Decoherence_Factor}), we propose the following model, with only two free parameters:
\begin{equation}\label{eq:Adiabatic_Ringdown}
    \begin{aligned}
        \Psi =& \Re\Bigl(A(v) e^{i \omega(v) v}\Bigr) \, , \\
        A(v) =& \Tilde{A} \Bigl[ 1 + \Tilde{Q} \frac{\delta m(v)}{m_2-m_1}\Bigr] \, , \quad \omega(v) = \frac{m_1\omega_{220}}{m(v)} \, , 
    \end{aligned}
\end{equation}
this is, the (complex) amplitude interpolates between some initial value $\Tilde{A}$ (which contains the two free parameters of the model), and a final amplitude given by $\Tilde{Q} \Tilde{A}$, and the ringdown frequency is equal to $m_1\omega_{220} / m(v)$. In the above, $\delta m(v) = m(v)-m_1$, and $m(v)$ is the time--varying mass, and $\omega_{220}$ is the (dimensionfull, complex) ringdown frequency of the Schwarzschild fundamental mode with mass $m_1$. We refer to this model as the \emph{dynamical ringdown} (DR) model, to distinguish it from the usual (linear) ringdown templates with constant amplitudes and frequencies. At this stage, this model is only adapted to a single mode. However, extending it to contain overtones and capture their time dependence, as well as the AIME effect, is possible, but we leave that for future explorations.

In order to compare this model with the usual damped sinusoids, we consider a situation that ressembles somewhat more closely what could happen after merger. We consider a reasonably mild accretion process, $m_2/m_1 = 1.05$. The accretion occurs ``shortly'' after the peak of the waveform, since we set $v_1 \sim 27$, with a timescale comparable to the decay rate of the fundamental mode, $\tau = 10m_1$. We inject our numerical waveforms into Gaussian white noise, defining at each timestep $t_j$
\begin{equation}
    \Psi_{\rm injection}[t_j] = \Psi_{\rm num}[t_j] + \varepsilon[t_j] \, , \quad \varepsilon \sim \mathcal{N}(0, \sigma = 10^{-4}) \, .
\end{equation}
We fit the waveform with both a damped sinusoids model containing one single damped sinusoid (we refer to this as the DS model), and with the DR model defined above. Comparing the performance of both models using the mismatch would not be completely fair, since they also contain different number of parameters. For this reason, we compute the Bayes factor between both models, defined as 
\begin{equation}
    \mathcal{B}_{\rm DR/DS} = \frac{p(d | \mathrm{DR})}{p(d | \mathrm{DS})} \, ,
\end{equation}
where $p(d | \mathrm{DR (DS)} )$ is the evidence for the model DR (DS) when observing the data $d$. We compute the evidence (and also estimate the parameters in the model) through a Bayesian method, sampling from the posterior distribution using the nested sampling algorithm~\cite{Skilling:2006gxv} using ellipsoidal bounding~\cite{mukherjeeNestedSamplingAlgorithm2006} as implemented in the \texttt{NestedSamplers.jl} package~\cite{NestedSamplersJL}. Our priors are uninformative uniform distributions in the ranges $\ln A \in [-5, 2]$~\footnote{For efficiency we sample on $\ln A$ rather than on the amplitude itself.}, $f \in [0, 1]$, $\tau \in [1, 20]$ and $\phi \in [0, 2\pi]$. We sample using $64$ live points, achieving fast convergence. 
Our code for the inference is made available through the \texttt{VaidyaPT.jl} repository~\cite{VaidyaPT}.

\begin{figure}
    \centering
    \includegraphics[width = \columnwidth]{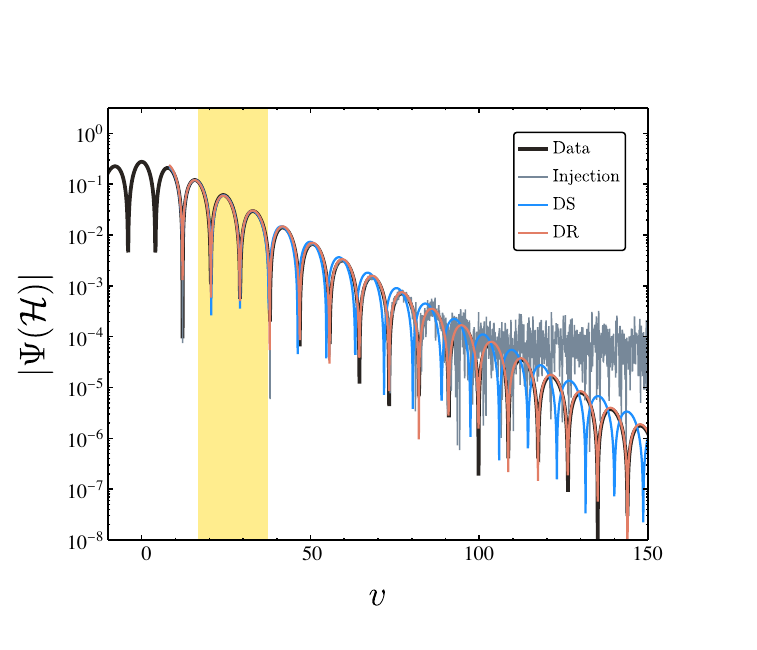}
    \caption{Numerical waveform (black) for a BH accreting $5\%$ of its initial mass $m_2/m_1 = 1.05$ with $v_1 = 40$ and $\tau = 10$, as extracted at the horizon. The initial data is given by a Gaussian profile with $(A_\Psi, v_\Psi, \sigma_\Psi) = (1, 10, 2)$. Overlapped, we show the injected waveform (including Gaussian white noise at a level of $10^{-4}$, and the reconstructed waveforms using the damped sinusoids (DS) model (in blue) and the dynamical ringdown (DR) model (in red). We use as starting and endtime of the fit $v_{\rm start} = 10$ and $v_{\rm end} = 150$. The yellow band represents the times at which spacetime is being highly dynamical. }   \label{fig:Bayesian_Reconstruction}
\end{figure}
The reconstructed waveform for the different models, compared to the injected waveform (including noise) as well as the numerical waveform, is shown in Fig.~\ref{fig:Bayesian_Reconstruction}. A simple visual inspection shows that the DR model captures much better the dynamical evolution of the waveform, even in the presence of noise. While the DS model is not able to capture the change in, e.g., the frequency and damping time of the ringdown due to accretion, this is perfectly captured by the DR model. Moreover, the amplitude and phase is consistent all throughout the evolution. The Bayes factor further supports this, obtaining a value of 
\begin{equation}
    \log \mathcal{B}_{DR/DS} \sim 3\times 10^{4} \, .
\end{equation}
This is very strong quantitative evidence, beyond the qualitative evidence provided by Fig.~\ref{fig:Bayesian_Reconstruction}, that in the presence of accretion a model containing varying amplitudes and frequencies is preferred with respect to a damped sinusoids (with fixed amplitudes and frequencies). 

%%%%%%%%%%%%%%%%%%%%%%%%%%%%%%%%%%%%%%%%
\section{Discussion}\label{sec:Discussion}
%%%%%%%%%%%%%%%%%%%%%%%%%%%%%%%%%%%%%%%%
We have studied gravitational perturbations of accreting spacetimes, namely a Vaidya geometry describing a BH growing through accretion of null radiation fields. This setup mimics that of a highly dynamical spacetime -- like that resulting from the coalescence of equal-mass BHs. Our interest in this problem is to complete our knowledge of the relaxation stage of BHs, and to understand whether one can extend a ringdown description based on (dynamical) damped sinusoids to earlier stages of the coalescence.

We showed that axial matter perturbations can be freely specified, and that they in turn source gravitational fluctuations that relax in a ringdown at late times. However, for the setups we considered, with null radiation being accreted, we find no evidence of other effects, such as echoes, that can arise in some situations due to the coupling to matter.

Our results indicate that
accreting BHs have quasi-characteristic modes, which are now a function of their mass, as might be expected for mass-changing spacetimes. Their ringdown process is qualitatively described by damped sinusoids, but with varying frequencies and amplitudes. By coupling these to the BH mass we provide a template, which might be useful in interpreting results from full nonlinear simulations of BH spacetimes. Another interesting question regards the behaviour of fluctuations when the accretion happens very slowly. For the timescales considered here we find that the accretion timescale does not seem to play a significant role, but a careful analysis of the adiabatic limit is left for future explorations. 

Breaking down the stationarity assumption of the background metric opens the way for new physics, of interest in realistic astrophysical scenarios. The framework developed here can be a starting point to study the gravitational dynamics in the presence of accretion in several configurations, such as its imprint on the inspiral of small bodies, or extending our findings to include a (time--varying) BH spin. 

%%%%%%%%%%%%%%%%%%%%%%%%%%%%%%%%%%%%%%%%%%%%%%%%%%%%%%% 
\section*{Acknowledgements}
%%%%%%%%%%%%%%%%%%%%%%%%%%%%%%%%%%%%%%%%%%%%%%%%%%%%%%
%
We thank Laura Sberna for providing the code used in Ref.~\cite{Green:2022htq} to compute the bilinear form on QNMs and comments on the manuscript.
J.R--Y. is indebted to Gregorio Carullo for guidance on the topics of data analysis and ringdown modelling, for stimulating conversations and detailed comments on the manuscript. 
We acknowledge fruitful discussions with Enrico Cannizzaro, Gregorio Carullo, Francisco Duque, Stephen Green, Michele Lenzi, Lionel London, Rodrigo Panosso--Macedo, Carlos Sopuerta and Rodrigo Vicente.
We also thank Jamie Bamber and Lionel London for valuable comments on the manuscript.
V.C.\ is a Villum Investigator and a DNRF Chair.  
The authors acknowledge support from the Villum Investigator program supported by VILLUM Foundation (grant no. VIL37766) and the DNRF Chair program (grant no. DNRF162), by the Danish Research Foundation.
V.C. acknowledges financial support provided under the European Union’s H2020 ERC Advanced Grant “Black holes: gravitational engines of discovery” grant agreement no. Gravitas–101052587. 
Views and opinions expressed are however those of the author only and do not necessarily reflect those of the European Union or the European Research Council. Neither the European Union nor the granting authority can be held responsible for them.
This project has received funding from the European Union's Horizon 2020 research and innovation programme under the Marie Sklodowska-Curie grant agreement No 101007855 and No 101007855.

\textbf{\emph{Software.}} The mansucript content has been derived using publicly available software: \texttt{DifferentialEquations.jl}, \texttt{Interpolations.jl}, \texttt{LineSearches.jl}, \texttt{NestedSamplers.jl}, \texttt{Noise.jl}, \texttt{Optim.jl}, and \texttt{Roots.jl} ~\cite{rackauckas2017differentialequations, mogensen2018optim, NestedSamplersJL}. This work makes use of the Black Hole Perturbation Toolkit~\cite{BHPToolkit}.

%%%%%%%%%%%%%%%%%%%%%%%%%%%%%%%%%%%%%%%%%%%%%%%%%%%%%%
%\clearpage
%%%%%%%%%%%%%%%%%%%%%%%%%%%%%%%%%%%%%%%%%%%%%%%%%%%%%%

\appendix

%%%%%%%%%%%%%%%%%%%%%%%%%%%%%%%%%%%%%%%%%%%%%%%%%%%%%%%%%%%%%%%%%%%%%%%%%
\section{Bilinear forms in the space of fluctuations}\label{App:Bilinear}
%%%%%%%%%%%%%%%%%%%%%%%%%%%%%%%%%%%%%%%%%%%%%%%%%%%%%%%%%%%%%%%%%%%%%%%%%
A useful tool that we make use of in Sec.~\ref{sec:Heuristics} is the bilinear form introduced in~\cite{Green:2022htq}, allowing to define a notion of orthogonality in the space of solutions of the Teukolsky equation. Here we review the construction of this bilinear form, as well as its basic properties. In the following, we work in the Teukolsky formalism but restricting to non--rotating BHs, i.e., the following can be recovered as the spinless limit of the general definition in the Kerr spacetime.

\begin{figure}
    \centering
    \includegraphics[width = 0.9\columnwidth]{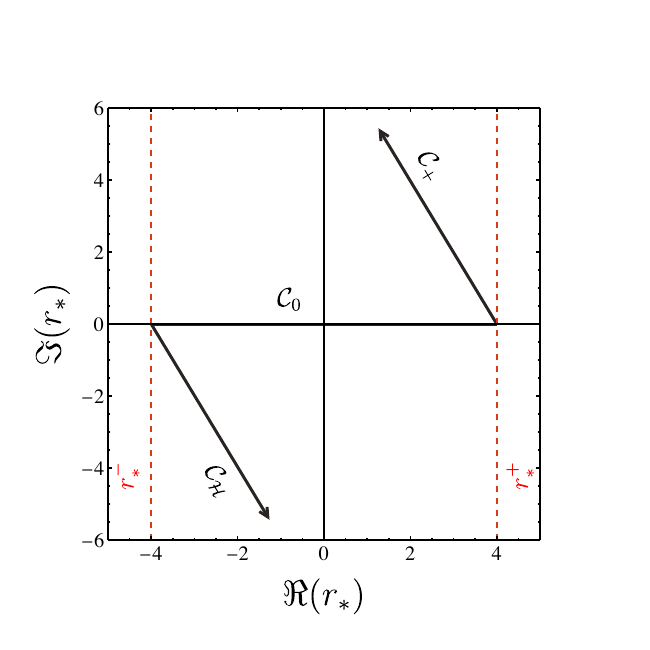}
    \caption{Integration contour $\mathcal{C}$ used to construct the bilinear form~\eqref{eq:Kinnersley-bilinearRAD}. Here, $r_*^{(-/+)}$ are the regularization parameters that control where is the contour modified. Convergence of this deformation scheme is shown extensively in~\cite{Sberna:2021eui, Green:2022htq}. }
    \label{fig:Integration_Contour}
\end{figure}

A fluctuation of the Weyl scalar $\psi_{0}$ is governed by a second order linear PDE 
\begin{equation}
    \mathcal{O}(\delta \psi_{0})=0\, ,
\end{equation}
where $\mathcal{O}(\cdot)$ is Teukolsky's operator \cite{Teukolsky:1973ha,Sberna:2021eui}. Associated to $\mathcal{O(\cdot)}$ there is an adjoint operator $\mathcal{O}^{\dagger}(\cdot)$ acting on functions\footnote{Working in the NP or GHP formalisms introduces further redundancies in the description of the gravitational field, namely the choice of frame, and a systematic approach to deal with those is that of principal fibre bundles. The idea is that in type $D$ spaces the Lorentz frame bundle can be reduced significantly by restricting to frames aligned with the (globally defined) principal null directions, and GHP scalars can be seen as sections in the associated vector bundles. This allows one to work in a language where it becomes manifest that equations like \eqref{Adj} are frame independent, a fact that is not obvious from the perspective that GHP scalars are just functions, as we do in our discussion.} which is uniquely defined by the equation \cite{Wald:1978vm}
\begin{equation}\label{Adj}
    \mathcal{O}^{\dagger}(\Upsilon)\tilde{\Upsilon}-\Upsilon\mathcal{O}(\tilde{\Upsilon})=\nabla^{\mu}\pi_{\mu}[\tilde{\Upsilon},\Upsilon]
\end{equation}
where $\pi_{\mu}[\tilde{\Upsilon},\Upsilon]$ is a local functional 1-form of any two functions $\tilde{\Upsilon}$ and $\Upsilon$ (both $\mathcal{O}^{\dagger}$ and $\pi_{\mu}[\tilde{\Upsilon},\Upsilon]$ follow by direct computation from Teukolsky's operator $\mathcal{O}$~\cite{Wald:1978vm}). If the equations $\mathcal{O}(\tilde{\Upsilon})= \mathcal{O}^{\dagger}(\Upsilon)=0$ hold (equivalently $\tilde{\Upsilon}\in\mathrm{ker}\mathcal{O}$ and $\Upsilon\in\mathrm{ker}\mathcal{O}^{\dagger}$), then $\star \pi[\tilde{\Upsilon},\Upsilon]$ is closed so its integral over any codimension-1 surface $\Sigma$, is invariant under local deformations $\Sigma'$ that keep the corners fixed, that is,
\begin{equation}\label{integral}
    \int_{\Sigma}\star \pi[\tilde{\Upsilon},\Upsilon]=\int_{\Sigma'}\star \pi[\tilde{\Upsilon},\Upsilon]\, .
\end{equation}
In particular, this means that if $\Sigma_{t}$ is a 1-parameter family of spacelike hypersurfaces with common corners, then \eqref{integral} evaluated on $\Sigma_{t}$ does not depend on $t$. That is, the integral \eqref{integral} is conserved in time. If, in addition, there is an operator $\mathcal{C}(\cdot)$ with the property $\mathcal{C}\Upsilon\in \text{ker}\mathcal{O}$ for all $\Upsilon\in\text{ker}\mathcal{O}^{\dagger}$ (i.e. mapping solutions of $\mathcal{O}^{\dagger}$ into solutions of $\mathcal{O}$), then
\begin{equation}\label{bilin}
    \ScalarProduct{\Upsilon_1}{\Upsilon_2}\equiv\int_{\Sigma}\star \pi[\mathcal{C}\Upsilon_{1},\Upsilon_{2}]
\end{equation}
gives rise to a bilinear form in the space of solutions $\text{ker}\mathcal{O}^{\dagger}$ that is conserved in time. Notice that the latter corresponds to the space of solutions with spin $s=-2$ (i.e. solutions for $\delta \psi_{4}$ instead of $\delta \psi_{0}$). The authors in \cite{Green:2022htq} considered a bilinear form \eqref{bilin} arising from the isometry $(t,\phi)\mapsto (-t,-\phi)$ present in Kerr's solution, and chose $\Sigma$ as a slice of constant Boyer-Lindquist (BL) time that extends from the bifurcation surface to spatial infinity. Explicitly, in the case of a non-rotating BH, in Schwarzschild coordinates and relative to a Kinnersley frame the product reads 
\begin{widetext}
\begin{align}\label{eq:Kinnersley-bilinear}
  \ScalarProduct{\Upsilon_1}{\Upsilon_2} 
  = m_{2}^{4/3} %\Bigg\{
  \int_\Sigma \dd  r \, \dd\theta \dd\phi\, \frac{\sin\theta}{r^{4}f^2} \left\{
  \Upsilon_1\Big|_{\substack{t\to-t \\ \phi\to-\phi}} \left(  \frac{r^{2}}{2f}\partial_t - 2 \left( r - \frac{m_{2}}{2f}\right) \right) \Upsilon_2 
    + \Upsilon_2 \left[\left(  \frac{r^{2}}{2f}\partial_t  - 2 \left( r -  \frac{m_{2}}{2f} \right) \right) \Upsilon_1\right]_{\substack{t\to-t \\ \phi\to-\phi}}
   \right\},
   \end{align}
\end{widetext}   
where $2f=1-2M_{2}/r$~\footnote{Notice that this is not the usual convention, however this is consistent with our convention for the Vaidya background, c.f. Eq.~\eqref{linel}.}. This bilinear is well defined and satisfies a number of desirable properties for solutions with compact support on $\Sigma$. This is, however, not the case of QNMs,
\begin{equation}\label{eq:modes}
 {}_s\Upsilon_{l \omega} = e^{-i\omega t} \Rh(r) \Sh(\theta),
\end{equation}
where we can restrict ourselves to axisymmetric modes, since we do not consider rotating BHs. Indeed, the spatial wavefunctions $\Rh(r)$ diverge both at the bifurcation surface and at spatial infinity. Specialised to modes of the form \eqref{eq:modes} (with $s=-2$) and using the orthogonality properties of the spherical harmonics $\Sh(\theta)$, the bilinear \eqref{eq:Kinnersley-bilinear} reads  
\begin{widetext}
\begin{equation}\label{eq:Kinnersley-bilinearRAD}
\ScalarProduct{\Upsilon_{l_{1}\omega_{1}}}{\Upsilon_{l_{2} \omega_{2}}}=2\pi m_{2}^{4/3}\delta_{l_{1}l_{2}}e^{-i(\omega_{2}-\omega_{1})t}\int \frac{dr_{\star}}{r^4f^{2}}R_{1}(r)R_{2}(r)\left(-i r^{2}\left(\omega_{1}+\omega_{2}\right)-4\left(r-3m_{2}\right)\right) \, , 
\end{equation}
\end{widetext}  
where we beware the reader that Eq. (55) in~\cite{Green:2022htq} contained some small typos. The above expression diverges if taken from $r_{\star }=-\infty$ to $r_{\star}=\infty$, where $dr_{\star}=dr/2f$ is the usual tortoise coordinate. In order to render this integral finite, the authors in \cite{Green:2022htq} proposed promoting $r_{\star}$ into a complex coordinate and performing the integral along a contour consisting of three pieces $\mathcal{C} = \mathcal{C}_\mathcal{H} \cup \mathcal{C}_0 \cup \mathcal{C}_+$, satisfying
\begin{equation}\label{eq:contour}
 \begin{cases}
   r_*(u,\epsilon)=u
   &\text{for $r_{*}^{-} < r_* <r_{*}^{+}$}\ \ \ (\mathcal{C}_0 )\\
   \arg r_*(u,\epsilon) \to \pi - \epsilon  & \text{for $r_* \to \infty$}\ \ \ (\mathcal{C}_+ ) \\
   \arg r_*(u,\epsilon) \to -\epsilon &\text{for $r_* \to -\infty$} \ \ \ (\mathcal{C}_{\mathcal{H}} ),
 \end{cases}
\end{equation}
where $r^{-}_{*}<0<r^{+}_{*}$ and $\epsilon>0$ are arbitrary parameters (see  Fig.~\ref{fig:Integration_Contour} for an example). 

As shown in~\cite{Green:2022htq}, with this prescription QNMs with $\omega_{1}\ne\omega_{2}$ are orthogonal relative to the bilinear~\eqref{eq:Kinnersley-bilinearRAD}. Moreover, if a function $\Upsilon$ admits a decomposition in QNMs $\Upsilon_{\ell m n}$ (where $n$ labels the overtones), 
\begin{equation}
    \Upsilon = \sum_{\ell m n} c_{\ell m n} \Upsilon_{\ell m n} \, , 
\end{equation}
then the coefficients
\begin{equation}
    c_{\ell m n} = \frac{\ScalarProduct{\Upsilon}{\Upsilon_{\ell m n}}}{\ScalarProduct{\Upsilon_{\ell m n}}{\Upsilon_{\ell m n}}} \, ,
\end{equation}
coincide with the QNM excitation coefficients as defined in the usual Laplace transform approach~\cite{Campanelli:1997un,Berti:2006wq}.

%%%%%%%%%%%%%%%%%%%%%%%%%%%%%%%%%%%%%%%%%
\section{Perturbation theory on spacetimes with generic matter and application to Vaidya's spacetime}\label{Ap1}
%%%%%%%%%%%%%%%%%%%%%%%%%%%%%%%%%%%%%%%%%%%
Perturbation theory on spherically-symmetric backgrounds simplifies due to the well-known decoupling between axial and polar sectors of the fluctuations. However, if the background is not vacuum gravitational and matter perturbations couple at first order, and this makes the analysis considerably more involved.  
We shall review briefly the formalism introduced in~\cite{Pereniguez:2023wxf}, which builds upon~\cite{Kodama:2000fa, Ishibashi:2003ap, Kodama:2003jz, Kodama:2003kk, Martel:2005ir, Chaverra:2012bh}, referring the interested reader to Ref.~\cite{Pereniguez:2023wxf} for details, and then apply it to the case of Vaidya spacetimes.

The basic idea is to consider background spacetimes whose metric and energy-momentum tensor have the general form
\begin{equation}
\begin{aligned}\label{GenlinelA}
ds^{2}&=g_{ab}(y)dy^{a}dy^{b}+r^{2}(y)\Omega_{AB}(z)dz^{A}dz^{B}\,, \\
T&=T_{ab}(y)dy^{a}dy^{b}+r^{2}(y)\mathcal{T}(y)\Omega_{AB}dz^{A}dz^{B}\, ,
\end{aligned}
\end{equation}
where the spacetime manifold has structure $M=\mathcal{N}^{2}\times \mathbb{S}^{2}$. Here, $g_{ab}(y)$, $T_{ab}(y)$, $r^{2}(y)$ and $\mathcal{T}(y)$ are a Lorentzian metric, a symmetric tensor, and two functions in $\mathcal{N}^{2}$, which is a 2-dimensional manifold parameterised by the coordinates $y^{a}$ (with $a=1,2$). The coordinates $z^{A}$ (with $A=3,4$) parameterise the unit round 2-sphere $\mathbb{S}^{2}$ with metric $\Omega_{AB}(z)$ (no assumption is made about the choice of neither $y^{a}$ nor $z^{A}$). In our background the line element is \eqref{linel}, while the energy-momentum tensor is 
\begin{equation}
T=\Phi K_{\mu}K_{\nu}dx^{\mu} dx^{\nu}=\Phi dv^{2}\, ,
\end{equation}
and therefore it falls in the general class of spacetimes \eqref{GenlinelA}. The metric and energy-momentum fluctuations, $h_{\mu\nu}$ and $\delta T_{\mu\nu}$, can be expanded in tensor harmonics as
\begin{align}\label{hdec}\notag
h=&\ h^{\ell}_{ab}(y)Y^{\ell}dy^{a}dy^{b}+2\left[h^{\ell}_{a}(y)Z^{\ell}_{A}+j^{\ell}_{a}(y)X^{\ell}_{A}\right]dy^{a}dz^{A} \\
&+\left[j^{\ell} (y)W^{\ell}_{AB}+k^{\ell} (y)U^{\ell}_{AB}+m^{\ell}(y)V^{\ell}_{AB}\right]dz^{A}dz^{B}\, , \notag\\ \\ \notag \\ \notag
\delta T=&\ \theta^{\ell}_{ab}(y)Y^{\ell}dy^{a}dy^{b}+2\left[\theta^{\ell}_{a}(y)Z^{\ell}_{A}+\rho^{\ell}_{a}(y)X^{\ell}_{A}\right]dy^{a}dz^{A} \notag \\% \label{Tdec}
&+\left[\rho^{\ell}(y) W^{\ell}_{AB}+\theta^{\ell}(y) U^{\ell}_{AB}+\sigma^{\ell}(y)V^{\ell}_{AB}\right]dz^{A}dz^{B}\, ,
\end{align}
where $Y^{\ell},Z^{\ell}_{A},U_{AB}^{\ell},V_{AB}^{\ell}$ and $X^{\ell}_{A},W^{\ell}_{AB}$ are the even and odd spherical tensor harmonics \cite{Pereniguez:2023wxf}, respectively, labelled by the harmonic indices $\ell=(l,m)$ and summation over repeated $\ell$'s is assumed (we may omit writing $\ell$ from now on). 
The so-called even and odd (equivalently polar and axial) sectors of the fluctuation consist of their components relative to the even and odd spherical harmonics, respectively (e.g.~$h^{\ell}_{ab}(y),h^{\ell}_{a}(y),k^{\ell} (y),m^{\ell}(y)$ form the even sector of $h_{\mu\nu}$ while $j^{\ell}_{a}(y),j^{\ell}(y)$ form the odd one). The linearised Einstein's equations can be written for these variables, and one finds a decoupling of sectors. It is also convenient to work with gauge-invariant variables. A gauge transformation acts on $h_{\mu\nu}$ and $\delta T_{\mu\nu}$ as
\begin{align}
h_{\mu\nu}&\mapsto h_{\mu\nu}-\pounds_{\xi}g_{\mu\nu}\, , \\ \notag \\
\delta T_{\mu\nu}&\mapsto \delta T_{\mu\nu}-\pounds_{\xi}T_{\mu\nu}\, ,
\end{align} 
where $\xi^{\mu}$ is a vector field. Then, it is easy to check that the fluctuation-dependent vector field $\eta[h]=\eta_{a}^{\ell}[h]Y^{\ell} dy^{a}+\left(\eta^{\ell}[h]Z^{\ell}_{A}+\upsilon^{\ell}[h]X^{\ell}_{A}\right)dz^{A}$, with 
\begin{align}\label{vec1}
\eta_{a}^{\ell}[h]&\equiv -h^{\ell}_{a}+\frac{r^{2}}{2}\nabla_{a}\left(\frac{m^{\ell}}{r^{2}}\right)\, ,\\ \label{vec2}
\eta^{\ell}[h]&\equiv -\frac{m^{\ell}}{2}\, ,\quad%\\\label{vec3}
\upsilon^{\ell}[h]\equiv -\frac{j^{\ell}}{2}\, ,
\end{align}
transforms as
\begin{equation}
\eta_{\mu}[h] \mapsto \eta_{\mu}[h]+\xi_{\mu}\,.
\end{equation}
Thus, the variables 
\begin{align}\label{htilde}
\tilde{h}&\equiv\left(h_{\mu\nu}+\pounds_{\eta}g_{\mu\nu}\right)dx^{\mu}dx^{\nu}\, , \\
\tilde{\theta}&\equiv\left(\delta T_{\mu\nu}+\pounds_{\eta}T_{\mu\nu}\right)dx^{\mu}dx^{\nu}\, ,
\end{align}
are manifestly gauge-invariant, and we shall work in terms of their harmonic components, denoted \footnote{The multipoles $l=0,1$ need to be treated separately~\cite{Pereniguez:2023wxf}.}
\begin{align}%\label{htildeCOMPS}
\tilde{h}=&\tilde{h}^{\ell}_{ab}Y^{\ell}dy^{a}dy^{b}+2\tilde{j}^{\ell}_{a}X^{\ell}_{A}dy^{a}dz^{A}+\tilde{k}^{\ell}U^{\ell}_{AB}dz^{A}dz^{B}\, ,\notag \\ 
\tilde{\theta}=&\tilde{\theta}^{\ell}_{ab}Y^{\ell}dy^{a}dy^{b}+2\left[\tilde{\theta}^{\ell}_{a}Z^{\ell}_{A}+\tilde{\rho}^{\ell}_{a}X^{\ell}_{A}\right]dy^{a}dz^{A} \notag\\
&+\left[\tilde{\rho}^{\ell}W^{\ell}_{AB}+\tilde{\theta}^{\ell}U^{\ell}_{AB}+ \tilde{\sigma}^{\ell}V^{\ell}_{AB}\right]dz^{A}dz^{B}\, .  \label{TtildeCOMPS}
\end{align}
The linearised Einstein's equations in terms of $\tilde{h}_{\mu\nu}$ and $\tilde{\theta}_{\mu\nu}$ can be found in \cite{Pereniguez:2023wxf}, together with the conservation laws of the energy-momentum tensor. 

Using the framework above, the only remaining task is to write the gauge-invariant energy-momentum tensor fluctuation $\tilde{\theta}_{\mu\nu}$ in terms of our particular matter model. Here we are considering null dust, which is described by a density function $\Phi$ and a null vector field $K^{\mu}$. The matter fluctuations are thus $\delta \Phi$ and $\delta K_{\mu}$, which we expand as 
\begin{align}
    \delta \Phi&=\mathcal{H} ^{\ell}Y^{\ell} \, , \\
    \delta K&=\mathsf{k}_{a}^{\ell}Y^{\ell} dy^{a}+\left(\mathsf{k}^{\ell}Z^{\ell}_{A}+\mathsf{v}^{\ell}X^{\ell}_{A}\right)dz^{A}\, .
\end{align}
Their gauge-invariant counterparts are
\begin{align}
\delta\Phi+\pounds_{\eta}\Phi&\equiv\tilde{\mathcal{H}}^{\ell}Y^{\ell}\, , \\
\delta K+ \pounds_{\eta}K&\equiv\tilde{\mathsf{k}}_{a}^{\ell}Y^{\ell} dy^{a}+\left(\tilde{\mathsf{k}}^{\ell}Z^{\ell}_{A}+\tilde{\mathsf{v}}^{\ell}X^{\ell}_{A}\right)dz^{A}\, ,
\end{align}
which explicitly read 
\begin{align}
\tilde{\mathcal{H}}^{\ell}&=\mathcal{H}^{\ell} +\eta^{\ell}_{a}\nabla^{a}\Phi \, ,\quad
\tilde{\mathsf{k}}^{\ell}_{a}=\mathsf{k}^{\ell}_{a}+\nabla_{a}\left(\eta^{\ell}_{b}K^{b}\right) \, , \\
\tilde{\mathsf{k}}^{\ell}&=\mathsf{k}^{\ell}+\eta^{\ell}_{b}K^{b}\, , \quad
\tilde{\mathsf{v}}^{\ell}=\mathsf{v}^{\ell}\, .
\end{align}
In terms of these, the gauge-invariant pieces of the energy-momentum tensor are
\begin{equation}
\begin{aligned}\label{emnulldust}
\tilde{\theta}^{\ell}_{ab}&=\tilde{\mathcal{H}}^{\ell}K_{a}K_{b}+2\Phi\tilde{\mathsf{k}}^{\ell}_{(a}K_{b)}\,,\\
\tilde{\theta}^{\ell}_{a}&=\Phi K_{a}\tilde{\mathsf{k}}^{\ell}\, ,\quad \tilde{\rho}^{\ell}_{a}=\Phi K_{a}\tilde{\mathsf{v}}^{\ell}\, .
\end{aligned}    
\end{equation}
while $\tilde{\rho}^{\ell}=\tilde{\theta}^{\ell}=\tilde{\sigma}^{\ell}=0$. 

The equations governing our fluctuation are the linearised Einstein's equations and the conservation law of the energy-momentum tensor given in \cite{Pereniguez:2023wxf}, where the components of the energy-momentum tensor are given by \eqref{emnulldust}. 

Next, we reduce such equations to decoupled wave equations in double-null coordinates, focusing on the axial sector which is the relevant one for this work. Consider first modes with $l>1$. The Einstein equations are \cite{Pereniguez:2023wxf}
\begin{equation} \label{einVVA}
\begin{aligned}
    r^{2}\nabla_{a}\left(r^{-2}\nabla^{a}\Omega\right)-\frac{\lambda^{2}-2}{r^2}\Omega&=r^{2}\varepsilon^{ab}K_{b}\nabla_{a}\left(\Phi \tilde{\mathsf{v}}\right)\, , \\
    \nabla_{a}\tilde{j}^{a}&=0\, ,
\end{aligned}
\end{equation}
where from now on $\nabla_{a}$ denotes the covariant derivative of the 2-dimensional metric $g_{ab}$, $\varepsilon_{ab}$ its natural volume-form, $\lambda^{2}=l(l+1)$, and the gravitational variable is 
\begin{equation}
    \Omega=-\frac{r^{4}}{2}\varepsilon^{ab}\nabla_{a}\left(\frac{\tilde{j}_{b}}{r^{2}}\right)\, .
\end{equation}
These equations are supplemented with (and actually imply) the conservation of the null-dust's energy-momentum tensor, which gives~\cite{Pereniguez:2023wxf}
\begin{equation}\label{consdustA}
    K^{a}\nabla_{a}\left(r^{2}\Phi \tilde{\mathsf{v}}\right)=0 \, .
\end{equation}
The first of Eqs.~\eqref{einVVA} and Eq.~\eqref{consdustA} are Eqs.~ \eqref{einVV} of the main text, with $\Omega=r \Psi$. Let us now consider the special odd mode, $l=1$. As shown in \cite{Pereniguez:2023wxf}, the solution is determined by a potential $\tau$, associated to the matter fluctuation, defined by
\begin{equation}\label{tau}
    \nabla_{a}\tau=r^{2}\varepsilon_{ab}\tilde{\rho}^{b}\, .
\end{equation}
Using the general solution for matter in \eqref{genMat}, equation \eqref{tau} gives
\begin{equation}
    \nabla_{a}\tau=-F(v)(dv)_{a} \ \ \longrightarrow \ \ \tau=-\int^{v}dv F(v)\, ,
\end{equation}
and the gravitational variable is simply given by
\begin{equation}
    \Omega(v)=-\int^{v}dv F(v)+\Omega_{0}\, ,
\end{equation}
where $\Omega_{0}$ is a constant. If the fluctuation $F(v)$ does not extend to future null infinity this mode asymptotically entails only a small change in angular momentum of the BH.

We conclude by relating $\Psi$ to the gravitational wave polarisations of the axial sector, $h_{+}$ and $h_{\times}$. For an in-going radiation field, and assuming that the in-falling wave is supported in a domain that does not extend to future null infinity (as is the case of the profile \eqref{eq:mass_profile}), by the same arguments of \cite{Martel:2005ir} the leading term of the axial metric fluctuation close to infinity in the radiation gauge is
\begin{equation}
    h^{\text{rad}}=r\sum_{\ell}\Psi^{\ell}_{\text{rad}}(u)W^{\ell}_{AB}dz^{A}dz^{B}\, ,
\end{equation}
where
\begin{equation}
    \Psi^{\ell}_{\text{rad}}(u)\equiv\frac{-4 }{\left(l-1\right)\left(l+2\right)}\Psi^{\ell}(u,r=\infty)\, ,
\end{equation}
and we recall that $u$ is the retarded time of the remnant BH. This, in turn, allows us to extract $h_{+}$ and $h_{\times}$ giving
\begin{equation}\label{eq:Psi_to_Strain}
    \begin{aligned}
    h_{+}&\equiv \frac{h^{\text{rad}}_{\theta\theta}}{r^{2}}=\frac{1}{r}\sum_{\ell}\Psi^{\ell}_{\text{rad}}(u)W^{\ell}_{\theta\theta}\, ,\\
     h_{\times}&\equiv \frac{h^{\text{rad}}_{\theta\phi}}{r^{2}\sin\theta}=\frac{1}{r\sin\theta}\sum_{\ell}\Psi^{\ell}_{\text{rad}}(u)W^{\ell}_{\theta\phi}\, .
\end{aligned}
\end{equation}

\bibliography{biblio.bib}
\end{document}